\documentclass{Interspeech2024}

\usepackage{makecell}
\usepackage{threeparttable}
\usepackage{stfloats}
\usepackage{multirow}
\usepackage{multicol}
\usepackage{cite}
\usepackage{array}
\usepackage{bbding}
\usepackage{tabularx}
\usepackage{graphicx}
\interspeechcameraready

\title{Perceiver-Prompt: Flexible Speaker Adaptation in Whisper for Chinese Disordered Speech Recognition}

\name[affiliation={1,4*}]{Yicong}{Jiang}
\name[affiliation={1,2*}]{Tianzi}{Wang}
\name[affiliation={1,4\dagger}]{Xurong}{Xie}
\name[affiliation={3}]{Juan}{Liu}
\name[affiliation={1}]{Wei}{Sun}
\name[affiliation={3}]{Nan}{Yan}
\name[affiliation={1,4}]{Hui}{Chen}
\name[affiliation={3}]{Lan}{Wang}
\name[affiliation={2}]{Xunying}{Liu}
\name[affiliation={1,4\dagger}]{Feng}{Tian}

\address{
  $^1$Institute of Software, Chinese Academy of Sciences, $^2$The Chinese University of Hong Kong\thanks{$*$Equal contribution}\\
  $^3$Shenzhen Institute of Advanced Technology, CAS,$^4$University of Chinese Academy of Sciences\thanks{$\dagger$ Corresponding authors}}
\email{\{xurong, tianfeng\}@iscas.ac.cn, jiangyicong231@mails.ucas.ac.cn, twang@se.cuhk.edu.hk}


\begin{document}

\maketitle

\begin{abstract}
    
    Disordered speech recognition profound implications for improving the quality of life for individuals afflicted with, for example, dysarthria. Dysarthric speech recognition encounters challenges including limited data, substantial dissimilarities between dysarthric and non-dysarthric speakers, and significant speaker variations stemming from the disorder. This paper introduces Perceiver-Prompt, a method for speaker adaptation that utilizes P-Tuning on the Whisper large-scale model. We first fine-tune Whisper using LoRA and then integrate a trainable Perceiver to generate fixed-length speaker prompts from variable-length inputs, to improve model recognition of Chinese dysarthric speech. Experimental results from our Chinese dysarthric speech dataset demonstrate consistent improvements in recognition performance with Perceiver-Prompt. Relative reduction up to 13.04\% in CER is obtained over the fine-tuned Whisper. 
\end{abstract}

\section{Introduction}

While significant strides have been made in conventional speech recognition, research into disordered speech recognition remains important due to its profound implications for improving the quality of life for individuals afflicted with, for example, dysarthria. Considering dysarthria as neuro-motor impairments resulting from neural damage affecting the muscular control involved in speech production, patients typically present with unclear articulation, variable speech rate, and disrupted speech rhythm\cite{kent2000research, rampello2016word}. Moreover, the scarcity of relevant datasets currently observed is attributable to the difficulty in collecting extensive data from individuals afflicted with dysarthria who experience physical disabilities and mobility constraints. These factors collectively contribute to (1) limited training data, (2) large mismatch between people with dysarthria and those without, and (3) large variations among speakers due to the effects of the disorder, posing significant challenges for dysarthric speech recognition.

To address these challenges, many studies have opted to train specialized ASR models using data from individuals with dysarthria. In \cite{joy2018improving}, researchers achieved the highest accuracy on the TORGO dysarthric speech database by adjusting acoustic model parameters, employing standardized cepstral features, and constructing complex DNN-HMM models. Other researchers proposed a method utilizing spectral-temporal deep features for both speech recognition and speaker adaptation\cite{geng2022spectro}, demonstrating its superiority over baseline methods through experimentation on the UASpeech corpus. A model adaptation approach for speaker-dependent dysarthric speech recognition systems\cite{9053725} first adapts to the general speaking style of multiple dysarthric speakers, followed by further adaptation to the target speaker. \cite{Shor_2019} achieves improved results by fine-tuning a specific subset of layers in the model, which involves significantly fewer parameters. LHUC(Learning Hidden Unit Contributions) is a parameter learning method for neural networks that enhances model performance by learning the importance weights of hidden units, particularly suitable for model adaptation and performance improvement\cite{swietojanski2016learning, Geng_2023, Liu_2021}. There are also methods based on i-vectors to model speaker and channel variability, thereby extracting speaker-specific information to enable the model to be applicable to disordered speech recognition tasks\cite{9900378, 5545402}.

With the emergence of pre-trained models such as Whisper\cite{whisper}, Hubert\cite{hubert}, Wav2Vec 2.0\cite{wav2vec}, among others, which leverage extensive data for pre-training, they can acquire universal speech representations from normal speech data. Such methodologies have the potential to compensate for the scarcity of dysarthric speech data. Fine-tuning these pre-trained models has become a recent trend. Both \cite{10097275} and \cite{geng23b_interspeech} have explored various approaches to incorporate domain-adapted self-supervised learning pre-trained models into speech recognition systems to address challenges encountered in recognizing dysarthric and elderly speech. Meanwhile, A few works\cite{baskar2022speaker} try to use x-vector and fMLLR for speaker adaptation on the self-supervised pre-trained Wav2vec2.0. However, speaker adaptation on large-scale pre-trained models, e.g. Whisper, remains a relatively underexplored area.

Ensuring pre-trained models exhibit robust performance on specific tasks constitutes a significant concern. To address this issue, numerous studies have yielded promising outcomes. LoRA\cite{lora} achieves rapid fine-tuning of large-scale language models by introducing low-rank parameters, thereby enhancing the model's efficiency and resource utilization in adapting to particular tasks or domains. Adapter Tuning\cite{adapter} involves adding small adapter layers to the parameters of pre-trained models, facilitating task-specific fine-tuning without altering the overall architecture. Particularly noteworthy is P-Tuning\cite{ptuning}, which incorporates trainable prompt embeddings optimized by a prompt encoder into inputs for improved performance, eliminating the need for manual prompt design.

P-Tuning is a method which is suitable for speaker adaptation in scenarios with limited speaker data. Its advantages lie in: (1) efficient utilization of limited data per speaker, (2) scalability to large-scale models with billions of parameters, and (3) the flexibility to capture different information with various configurations. Building upon this, we propose the Perceiver-Prompt for Speaker Adaptation on the Whisper large-scale speech model. The Perceiver-Prompt is a sequence of vectors derived by a P-tuning trained Perceiver using data from the same speaker. This approach aims to enhance the adaptation of the model to disordered speech recognition tasks, effectively leveraging data to address the variability and the scarcity of data. In order to enable the large-scale pre-trained model to compensate for the scarcity of disordered speech data, we employ LoRA to fine-tune the medium-sized Whisper model on our constructed dataset of disordered speech. Whereafter, a Perceiver is trained by P-tuning and serves as a speaker-dependent prompt generator of Whisper, using input from utterances of the same speaker. Subsequent evaluations are conducted on our dataset of Chinese dysarthric speech, where the proposed Perceiver-Prompt based speaker adaptation obtains consistent recognition performance improvements using various configurations.

The main contributions of this paper are summarized as follows: 
(1) To the best of our knowledge, this is the first work of applying P-Tuning for speaker adaptation on the Whisper large-scale speech model;
(2) Our proposed Perceiver-Prompt can handle variable-length inputs and produce fixed-length speaker prompts; 
(3) Experiments conduct on our dysarthric speech dataset shows that, our method achieves relative reduction up to 13.04\% in Character Error Rate (CER) over the fine-tuned Whisper; Particularly, on the speech with highest severity level, the relative CER reduction is up to 51.38\%;
(4) Further Investigation is conducted for the correlation between the effectiveness of our proposed Perceiver-Prompt and the ability to discriminate between speakers or dysarthria severity levels.

The rest of this paper is organized as follows. Section 2 introduces the large-scale pre-trained model Whisper and the foundational P-Tuning for speaker adaptation. Section 3 introduces a prompt encoder Perceiver and proposes a training method Perceiver-Prompt that combines Perceiver with P-Tuning. Section 4 presents the experimental methodology and results. Finally, conclusions are drawn in Section 5.

\section{Preliminary}
\subsection{Large-scale pre-trained model Whisper and LoRA}
Whisper\cite{whisper} is an automatic speech recognition (ASR) system trained on a large dataset of 680,000 hours of multilingual and multitask supervised data. It employs an end-to-end encoder-decoder architecture based on Transformer. Input audio is split into 30-second segments and converted into log-Mel spectrograms, while the decoder predicts text tokens with specialized tokens for tasks like language identification and translation to English. With about one-third of its data being non-English, Whisper is suitable for Mandarin recognition due to its multilingual capabilities\cite{whisper}.

Despite Whisper's extensive training and demonstrated accuracy in various scenarios, it struggles to generalize to disordered speech datasets due to their variability and low clarity. This can lead to recognition errors like unpredictable sentence terminations or misidentification of excessive nasality as speech\cite{ENDERBY2013273}. Fine-tuning on specific datasets is necessary to enhance Whisper's adaptability to dysarthric speech recognition tasks.

Considering computational costs and the effectiveness of fine-tuning, we opt to employ LoRA for fine-tuning Whisper on our dataset of disordered speech. LoRA is designed to encode the parameters of pre-trained models using a low-rank matrix, which does not incur additional inference time and is more amenable to optimization\cite{lora}. The process of fine-tuning using LoRA proceeds as follows:
\begin{align}
  W_0 + \Delta W = W_0 + BA, B \in \mathbb{R}^{d \times r}, A \in \mathbb{R}^{r \times k}
\end{align}
where $W_0$ represents the frozen parameters in the pre-trained model, $\Delta W$ denotes the parameters requiring fine-tuning, $B$ and $A$ are the up-projection matrix and down-projection matrix and $r \ll \min (d, k)$ respectively. During training, only $B$ and $A$ are trained. $r$ is empirically set as 8 during training and fixed throughout the experiments of this paper.

\subsection{P-tuning}
Integrating discrete or continuous prompts as additional inputs into language models can yield improved performance\cite{ptuning}. However, \cite{ptuning} indicates that the use of discrete prompts imposes high design requirements and improper utilization may lead to model instability, potentially limiting the exploitation of gradient descent in discrete space searches. Therefore, P-Tuning utilizes a separate Prompt Encoder to generate corresponding continuous Prompt Embeddings as additional inputs, achieving superior results. This design can be applied to Speaker Adaptation in speech recognition, utilizing the Prompt Encoder to encode certain speaker data into Prompt Embeddings representing Speaker Features as additional inputs to the model. In tasks involving disordered speech recognition, distinct Speaker Features may signify different articulatory disorder symptoms, aiding the model's adaptation to disordered speech recognition tasks. 

\section{Perceiver-Prompt}
\subsection{Employing Perceiver as a Prompt Encoder}
To implement P-Tuning for speaker adaptation in disordered speech recognition tasks, it is necessary to augment the Whisper with a Prompt Encoder to handle complex temporal speech data and vectorize its speaker characteristics to form the model's Prompt. Perceiver\cite{jaegle2021perceiver} emerges as an excellent choice for this purpose. The advantage of Perceiver lies in its ability to generate Prompts without making specific domain assumptions and handle input with arbitrary form. It generates fixed-length embedding vectors based on input, facilitating concatenation with input data without requiring data preprocessing. Additionally, another crucial factor is Perceiver's capability to generate speaker prompts online, enabling relatively efficient adaptation during inference. The Perceiver-based prompt Encoder is shown in figure \ref{fig:architecture} (left, green).

\subsection{Perceiver-Prompt for speaker adaptation}
\begin{figure}
    \centering
    \includegraphics[width=0.8\linewidth]{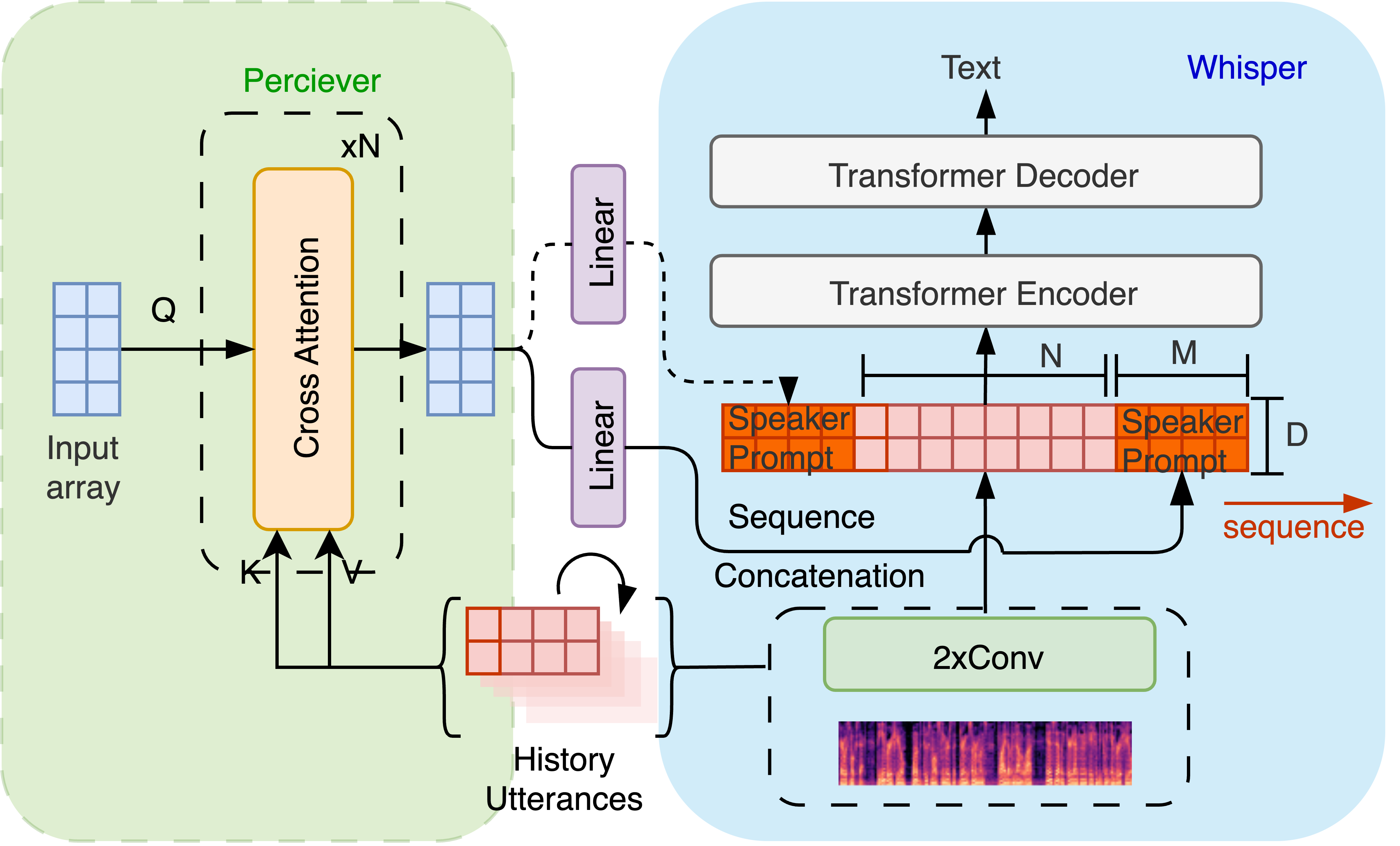}
    \setlength{\belowcaptionskip}{-0.5cm}
    \caption{The flexible concatenation method of Perceiver-Prompt (optionally including data from the same speaker).}
    \label{fig:architecture}
\end{figure}

For speaker adaptation, the proposed Perceiver-Prompt can be trained with P-tuning in a flexible way by employing various configurations. In general, the speaker prompt is a sequence of vectors generated by a Perceiver followed by a linear transform. The Perceiver input is an embedding sequence consisting of various number of history utterances from the same speaker on the adapted Whisper layer. The linear transform maps the speaker prompt to the same feature space with the input embeddings. Subsequently, the obtained speaker prompt is concatenated with the input embedding sequence at various positions to form a new embedding sequence for speaker adaptation. Such positions can be the beginning, end, or both sides of the sequence, as shown in figure \ref{fig:architecture}, right, blue. Finally, parameters of the Perceiver-Prompt is trained with P-tuning by fixing the parameters of Whisper. Taking the concatenation at the end of input as an example, this process can be represented as follows:
\begin{small}
\begin{equation}
  \begin{aligned}
      & \text{Prompt} = \text{Perceiver}(\mathbf{X_p}), \quad\text{Prompt} \in \mathbb{R}^{M, D}\\
      & \mathbf{X_{\text{new}}}=\text{Concat}(\mathbf{X},(\mathbf{W}(\text{Prompt})+\mathbf{b}))\\
      & \mathbf{X} \in \mathbb{R}^{N, D}, \mathbf{X_{\text{new}}} \in \mathbb{R}^{M+N, D} 
  \end{aligned}
\end{equation}
\end{small}
\noindent where $\mathbf{X}$ is the length-$N$ embedding sequence with dimensionality $D$ of a layer, $\mathbf{X_{new}}$ denotes the new length-$(M+N)$ embedding sequence with dimensionality $D$ of the layer, $\mathbf{X_p}=\text{Concat}(\mathbf{X},\mathbf{S_1},\mathbf{S_2},...,\mathbf{S_n})$ denotes the input of Perceiver, and [$\mathbf{S_1},...,\mathbf{S_n}$] represent the utterances from the same speaker utilized. For decoding, the speaker prompt capturing speaker information can be generated from the Perceiver-Prompt efficiently by using the history utterances from the target speaker as input, and then used by the Whisper for speaker adaptation. 

In practice, the Perceiver-Prompt can be positioned on the input log-mel features of Whisper, or positioned before the encoder blocks in Whisper. Moreover, additional information can also be employed as auxiliary supervision for Perceiver-Prompt training to assist the model in better adapting to disorder speech recognition tasks, such as speaker identity or disorder severity.

\section{Experiments}
\subsection{Experimental Setup}
\textbf{Dataset: }We constructed an dysarthric speech dataset following the approach outlined in \cite{liu2023audio}, comprising Chinese speech data collected from 64 patients with varying degrees of dysarthria (approximately 31 hours) and 20 healthy participants (approximately 18 hours). Collection tasks included reading tasks involving Chinese characters, words and sentences. Additionally, we gathered Frenchay Dysarthria Assessment (FDA) scale information from patients, segmented into four severity levels based on scores of \{135, 90, 70\}/145. The test set comprised speech data from 11 patients with varying degrees of dysarthria, while the remaining data from 53 patients and 20 healthy participants were used for training. In the subsequent experiments, we will evaluate the model performance under different levels of FDA severity and various speech tasks. In the following experiments, F.1 represents the highest severity level of articulatory disorders, while F.4 represents the least severe. Additionally, T.1 denotes Chinese character recognition tasks, T.2 denotes word recognition tasks, and T.3 denotes sentence recognition tasks.

\textbf{Whisper: }We chose Whisper-medium as the experimental base due to its combination of effective performance and relatively conservative computational resources. The ESPnet\cite{watanabe2018espnet} recipe provided us with a pre-configured setup for fine-tuning Whisper-medium using LoRA. Both Encoder and Decoder are stacked with 24 Residual Attention Blocks, each block consisting of multi-head Attention and a Feed-Forward Network. The fine-tuning objective of LoRA targets components within the Attention module such as "query", "key", "value", and "att.out", with a LoRA rank set to 8. In Whisper, all audio is resampled to 16,000 Hz and represented using 80-channel log amplitude mel-spectrograms computed with a 25-millisecond window and a 10-millisecond stride. 

\textbf{Perceiver: }We utilized the Perceiver provided by \cite{Eren_Coqui_TTS_2021} to generate a latent array of length 32, which represents speaker features. The feature dimension was set to 1024, matching the input of the Whisper Encoder blocks, with the number of heads in the internal multi-head attention set to 8.

\subsection{General Result Analysis}
While Whisper has been trained on a considerably large dataset and can achieve human-like recognition results in most tasks without the need for fine-tuning, it is not directly applicable to dysarthric speech recognition tasks. We directly performed inference using the pre-fine-tuned Whisper-medium and Whisper-large, obtaining CERs of 141.9\% and 87.9\%, respectively. The higher CER results from Whisper's difficulty in predicting sentence terminations on the dysarthric speech dataset before fine-tuning. Hence, all Whisper systems used later have been fine-tuned with LoRA using the dysarthric speech.

The state-of-the-art End-to-end Conformer\cite{conformer} and hybrid TDNN\cite{Peddinti2015ATD} systems without pre-training are employed for comparison with our approach. Meanwhile, i-vector, as an effective and widely used method for extracting speaker features, is also of interest for comparison. The result shows in table \ref{tb:table1} indicate that with the adaptation on Whisper by our proposed Perceiver-Prompt, the model achieved a reduction of 13.04\% relative (0.9\% absolute ) in CER compared to the baseline model, Whisper-medium, and outperformed the i-vector adapted Whisper as well as the Conformer and TDNN systems without pre-training. For speech samples with higher level of articulatory disorders (F.1, F.2), the proposed approach still demonstrates the best performance, particularly for the most severe F.1, with a significant relative CER reduction of 35.78\% relative (3.9\% absolute).
\begin{table}[htbp]
    \scriptsize
    \setlength{\abovecaptionskip}{0cm}
    \setlength{\belowcaptionskip}{0cm} 
    \caption{The performance of Conformer, TDNN, fine-tuned Whisper-medium, Whisper-medium with speaker adaptation using i-vectors (Whisper-iVector) and with the proposed \textbf{Perceiver-Prompt (Whisper-PP}) method by concatenating Speaker Prompt at the end of the input.}
    \centering
    \begin{threeparttable} 
    \setlength{\tabcolsep}{4pt}
    \begin{tabular*}{\linewidth}{@{}ccccccccc@{}}
        \toprule
        \multirow{2}{*}{Model} & \multicolumn{4}{c}{FDA-CER(\%)} & \multicolumn{3}{c}{TSK-CER(\%)} & \multirow{2}{*}{CER(\%)}   \\
        & F.1 & F.2 & F.3 & F.4 & T.1 & T.2 & T.3 &  \\
        \midrule
        Conformer                  & 33.1 & 43.8 & 6.3 & 1.8 & 9.6 & 14.9 & 14.5 & 12.9\\
        TDNN                       & 61.4 & 22.6 & 2.9 & 0.3 & 7.3 & 8.7 & 14.1 & 10.4\\
        Whisper-medium             & 10.9 & 18.8 & 5.7 & 1.4 & 6.1 & 7.0 & 7.4 & 6.9\\
        Whisper-iVector            & 13.5 & 20.7 & 6.2 & 1.4 & 7.9 & 8.9 & 6.2 & 7.6\\
        \bfseries Whisper-PP & \bfseries 7.0 & \bfseries 16.1 & \bfseries 5.4 & \bfseries 1.4 & \bfseries 5.3 & \bfseries 6.5 & \bfseries 6.2 & \bfseries 6.0 \\
        \bottomrule
    \end{tabular*}
    \end{threeparttable} 
    \label{tb:table1}
\end{table}

\vspace{-0.2cm}
\subsection{Different Configuration}
Perceiver-Prompt demonstrates promising results across various configurations due to its highly flexibility. We conducted experiments on the placement of Perceiver, concatenation positions with inputs, the number of historical speech instances used, the length of Speaker Prompt, and other configurations. The experimental outcomes, as presented in Table \ref{tb:table2}, showcase the superiority of Perceiver-Prompt in dysarthric speech recognition tasks. It adapts to different tasks by adjusting configurations; for instance, employing 5 historical utterances from the same speaker (Conf.8) to generate Speaker Prompt achieves the lowest 5.3\% CER for the most severe F.1 (with 51.38\% relative CER reduction over the unadapted Whisper). This might be attributed to the model's ability to learn continuous articulation characteristics of patients from historical information. Similarly, for sentence recognition tasks, Conf.2 yields the best performance with CER of 6.2\%. 

In summary, our Perceiver-Prompt demonstrates considerable flexibility in configurations and consistently outperforms the baseline with various configurations and achieves the lowest CER of 6.0\%. Specifically, the minimum word error rate is 5.3\%, 6.3\%, and 6.2\% for the Chinese character, word, and sentence recognition tasks, respectively.

\begin{figure}
    \centering \hspace{-2mm}
    \includegraphics[width=\linewidth]{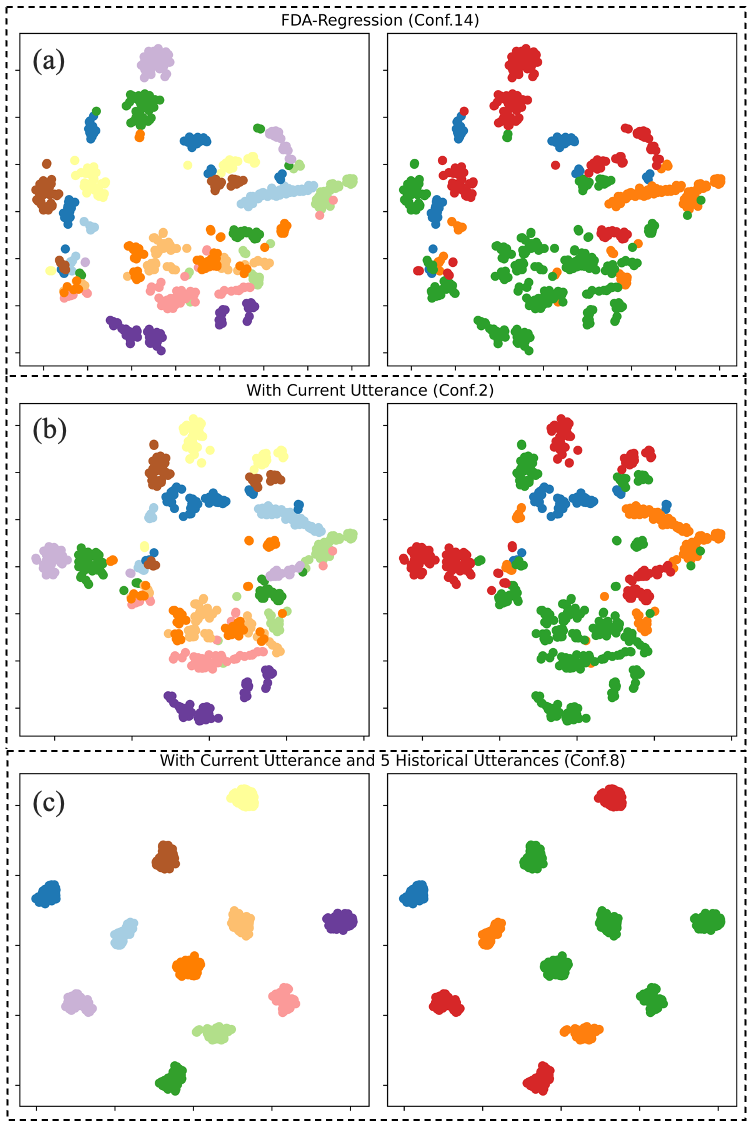}
    \setlength{\abovecaptionskip}{-0.3cm}
    \setlength{\belowcaptionskip}{-0.6cm}
    \caption{The three subplots (a), (b), and (c) respectively represent the t-SNE clustering results of Conf.14, Conf.2, and Conf.8. The left column represents t-SNE analysis for speakers, with different colors indicating different speakers. The right column represents t-SNE analysis for FDA severity levels, with different colors indicating different FDA severity levels.}
    \label{fig:tsne}
\end{figure}

\begin{table}[t]
    \small
    \caption{The performance of Perceiver-Prompt under various configurations and training methods. All configurations improve the model in better adapting to dysarthric speech recognition, and targeted adjustments in specific scenarios can lead to improved performance. ``self'' denotes the current utterance. ``Sto.'' refers to stochasticity select the data from the same speaker.}
    \setlength{\abovecaptionskip}{0cm}
    \setlength{\belowcaptionskip}{-0.3cm}
    \resizebox{\linewidth}{!}{
    \begin{tabular}{c|c|c|c|c|c|c|c|c|c|c|c|c|c}
        \hline
        \hline \rule{0pt}{8pt}
        \multirow{2}{*}{Conf.} & \multicolumn{2}{c|}{Position} & \multicolumn{2}{c|}{Input Utterances}  & \multirow{2}{*}{\makecell{Prompt\\length}} & \multicolumn{4}{c|}{FDA-CER(\%)} & \multicolumn{3}{c|}{TSK-CER(\%)} & \multirow{2}{*}{\makecell{CER\\(\%)}}  
        \\
        \cline{2-5}\cline{7-13}\rule{0pt}{8pt}
        & Concat & Layer & history & Sto. & & F.1 & F.2 & F.3 & F.4 & T.1 & T.2 & T.3 & \\
        \hline
        \hline \rule{0pt}{8pt}
        1 &  &   & self & - & 64 & 5.3 & 19.5 & 5.6 & 1.4 & 6.0 & 6.6 & 6.7 & 6.4\\
        \cline{1-1}\cline{4-5}\cline{6-14}\rule{0pt}{9pt}
        2 & End &  & self & - & 32 & 7.0 & 16.1 & 5.4 & 1.4 & 5.3 & 6.5 & 6.2 & \textbf{6.0}\\
        \cline{1-1}\cline{4-5}\cline{6-14}\rule{0pt}{9pt}
        3 &  &   & self & - & 16 & 13.1 & 16.8 & 5.6 & 1.4 & 5.8 & 6.4 & 7.8 & 6.7\\
        \cline{1-2}\cline{4-14}\rule{0pt}{9pt}
        4 & Beginning &  & self & - &  & 9.4 & 18.6 & 5.6 & 1.6 & 6.1 & 7.1 & 6.9 & 6.7 \\
        \cline{1-2}\cline{4-5}\cline{7-14}\rule{0pt}{9pt}
        5 & Both sides & \multirow{12}{*}{\centering \makecell{Before\\encoder\\blocks \\ \\ \\ \\ \\ \\ \\ \\ \\ \\} } & self & - &  & 5.3 & 18.8 & 5.7 & 1.4 & 6.0 & 6.7 & 6.4 & 6.4 \\
        \cline{1-2} \cline{4-5}\cline{7-14}\rule{0pt}{9pt}
        6 & \multirow{6}{*}{End} &  & self+1 &  & \multirow{10}{*}{\makecell{ 32 \\ \\ \\ \\ \\ \\  }} & 5.3 & 19.2 & 5.7 & 1.4 & 5.9 & 6.6 & 6.8 & 6.4 \\
        \cline{1-1}\cline{4-4}\cline{7-14}\rule{0pt}{9pt}
        7 &  &  & self+3 & \XSolidBrush &  & 7.0 & 18.3 & 5.7 & 1.5 & 6.1 & 7.0 & 6.4 & 6.5\\
        \cline{1-1}\cline{4-4}\cline{7-14}\rule{0pt}{9pt}
        8 &  &  & self+5 &  &  & 5.3 & 18.1 & 5.7 & 1.5 & 6.3 & 6.3 & 6.3 & 6.3\\
        \cline{1-1}\cline{4-5}\cline{7-14}\rule{0pt}{9pt}
        9 &  &  & self+1 &  &  & 12.3 & 19.2 & 5.4 & 1.5 & 5.8 & 7.0 & 8.0 & 6.9\\
        \cline{1-1}\cline{4-4}\cline{7-14}\rule{0pt}{9pt}
        10 &  &  & self+3 & \Checkmark &  & 9.9 & 19.2 & 5.7 & 1.5 & 6.3 & 7.0 & 7.4 & 6.9\\
        \cline{1-1}\cline{4-4}\cline{7-14}\rule{0pt}{9pt}
        11 &  &  & self+5 &  &  & 8.2 & 18.5 & 5.7 & 1.4 & 6.2 & 6.5 & 7.1 & 6.6\\
        \hline 
        \hline \rule{0pt}{9pt}
        12 & End & \makecell{log-mel} & self & - & 32 & 6.5 & 19.4 & 5.7 & 1.5 & 6.1 & 6.7 & 7.0 & 6.6\\
        \hline
        \hline
    \end{tabular}
    }
    \label{tb:table2}
\end{table}
\subsection{Joint training with additional information}
We attempted auxiliary supervised learning by incorporating FDA and Speaker identity as additional information to aid Perceiver-Prompt in adapting better to dysarthric speech recognition tasks. We primarily employed three methods of joint training utilizing MLP , where three auxiliary tasks employ different supervisions during Perceiver-Prompt training. (1) speaker classification for each speech utterance, (2) FDA score regression for each speech utterance, and (3) classification of FDA severity levels for each speech utterance, where normal speech was treated as a fifth class. The results of joint training using different approaches are presented in the table \ref{tb:table3}, indicating that joint training still yields favorable outcomes, with superior performance observed on the sentence recognition task (T.3) when leveraging FDA  regression. Particularly, it yields the best performance with a 6.0\% CER for T.3.
\begin{table}[hbp]
    \scriptsize
    \setlength{\abovecaptionskip}{0cm}
    \setlength{\belowcaptionskip}{-0.3cm} 
    \caption{Model performance under different joint training approaches. Configurations remain consistent: concatenation of a length-32 Speaker Prompt at the end of Encoder blocks' input, without utilizing any historical speech data.}
    \centering
    \begin{threeparttable} 
    \setlength{\tabcolsep}{4.5pt} 
    \begin{tabular*}{\linewidth}{@{}c|c|c|c|c|c|c|c|c|c@{}}
        \toprule
        \hline \rule{0pt}{8pt}
        \multirow{2}{*}{Conf.}  & \multicolumn{4}{c|}{FDA-CER(\%)} & \multicolumn{3}{c|}{TSK-CER(\%)} & \multirow{2}{*}{\makecell{CER\\(\%)}} & \multirow{2}{*}{\makecell{Joint \\ training}}  \\
        \cline{2-8}\rule{0pt}{10pt}
        & F.1 & F.2 & F.3 & F.4 & T.1 & T.2 & T.3 & \\
        \hline
        \hline \rule{0pt}{8pt}
        13 & 7.0 & 19.6 & 5.5 & 1.4 & 6.0 & 7.0 & 6.8 & 6.6 &\makecell{ Speaker-classify\\\& ASR}\\
        \cline{1-10}\rule{0pt}{10pt}
        14 & 7.2 & 17.2 & 5.7 & 1.4 & 6.1 & 6.9 & 6.0 & 6.3 &\makecell{FDA-regress \\\& ASR}\\
        \cline{1-10}\rule{0pt}{10pt}
        15 & 8.2 & 17.2 & 5.7 & 1.4 & 5.9 & 6.5 & 6.8 & 6.4 & \makecell{FDA-classify \\\& ASR}\\
        \hline
        \bottomrule
    \end{tabular*}
    \end{threeparttable} 
    \label{tb:table3}
\end{table}
\vspace{-0.3cm}
The Speaker Prompts are flattened and subjected to t-SNE analysis, as depicted in the figure \ref{fig:tsne}. It can be observed that the Speaker Prompts derived from Perceiver-Prompt can partly differentiate different speakers and various levels of FDA severity. To delve deeper into underlying patterns, Speaker Prompts from Conf.2 and Conf.8 were processed with different history, yielding results shown in the figure \ref{fig:tsne}. It is evident that incorporation of more historical information enhances speaker discrimination. However, this does not necessarily guarantee improved performance in dysarthric speech recognition tasks.

\vspace{-1mm}
\section{Conclusion}
\vspace{-1mm}
This paper proposes the Perceiver-Prompt, a speaker adaptive method based on P-tuning in Whisper for dysarthric speech recognition. By employing various configurations and techniques such as multi-task learning with auxiliary FDA score regression or multiple history utterances, this approach has the flexibility to produce adaptable prompt to accommodate varying dysarthric severity levels or different speech tasks. On our Chinese dysarthric speech dataset, the fine-tuned Whisper applying Perceiver-Prompt outperforms the Conformer, TDNN, and fine-tuned Whisper with/without using i-vector, by achieving a relative CER reduction up to 13.04\% across all configurations. To the best of our knowledge, this is the first work of applying P-Tuning for speaker adaptation on the Whisper largescale speech model.

\section{Acknowledgement}
This paper is supported by National Science and Technology Major Project (2022ZD0118002), Research Project of Institute of Software, Chinese Academy of Sciences (ISCAS-ZD-202401, ISCAS-JCMS-202306), National Natural Science Foundation of China (62106255, U23B2018 and 62271477 ), Youth Innovation Promotion Association CAS Grant (2023119), Shenzhen Peacock Team Project (KQTD20200820113106007), Hong Kong RGC GRF grant No.14200021, 14200218, 14200220, TRS T45-407/19N, Innovation \& Technology Fund grant No. ITS/254/19, ITS/218/21.

\bibliographystyle{IEEEtran}
\bibliography{interspeech2024}

\end{document}